\documentclass[useAMS,usenatbib]{mn2e}

\usepackage{epsfig}
\usepackage{amsmath}

\title{$\phi^2$ as Dark Matter}

\author[Matos et al.]
{Tonatiuh Matos $^{1,3}$\thanks{E-mail:tmatos@fis.cinvestav.mx},
Alberto V\'azquez-Gonz\'alez $^{1,3}$\thanks{E-mail:jvazquez@fis.cinvestav.mx}
and Juan Maga\~na$^{2,3}$\thanks{E-mail:jmagana@astroscu.unam.mx}\\
$^{1}$Departamento de F{\'\i}sica, Centro de Investigaci\'on y de Estudios Avanzados del
IPN, A.P. 14-740, 07000 M\'exico D.F., M\'exico.\\
$^{2}$Instituto de Astronom\'{\i}a, Universidad Nacional
Aut\'onoma de M\'exico, A.P. 70-543, C.P. 04510, D.F., M\'exico\\
$^{3}$Part of the Instituto Avanzado de Cosmolog\'ia (IAC) collaboration http://www.iac.edu.mx/}

\begin{document}

\date{Accepted xxxx Month xx. Received xxxx Month xx; in original form
2008 May 23}
\pagerange{\pageref{firstpage}--\pageref{lastpage}} \pubyear{2008}
\maketitle

\label{firstpage}

\begin{abstract}

In this paper we consider $\phi^2$ scalar field potential as a candidate to dark
matter. If it is an ultralight boson particle, it condensates like a Bose-Einstein
system at very early times and forms the basic structure of the Universe.
Real scalar fields collapse in equilibrium configurations 
which oscillate in space-time (oscillatons).The cosmological
behavior of the field equations are solved using
the dynamical system formalism. We use the current cosmological parameters as
constraints for the present value of the scalar field
and reproduce the cosmological predictions of the standard
$\Lambda$CDM model with this model. Therefore, scalar field dark matter
seems to be a good alternative to cold dark matter nature.
\end{abstract}


\section{Introduction}

Scalar fields are one of the most interesting and most mysterious
fields in theoretical physics. Fundamental scalar fields are needed
in all unification's theories, however, there are not
experimental evidence of its existence.
From the standard model of particles which needs the Higgs boson, until the
superstring theory which contains the dilaton, passing throught the
Kaluza-Klein and the Brans-Dicke theories or throught the
inflationary model, scalar fields are necessary fields. Doubtless,
if they exist, they have some features which make them very special.

The \textit{Scalar Field Dark Matter} (SFDM) model
paradigm has been constructed step by step. One of the
first suggestions that a (complex) scalar field could contribute
to structure formation of the Universe was given by \citet{press}
and \citet{madsen}. Nevertheless, complex scalar fields were used before as matter candidates as boson stars by \citet{ruffini} (for a recent introduction to boson stars, see for example \cite{paco06}). One of the first candidates to be scalar field dark matter is the axion, one of the solutions to the strong-CP problem in QCD (see an excellent review in \cite{Kolb}). Essentially, the axion is a scalar field with mass restricted by observations to $\sim10^{-5}$eV, which has its origin at $10^{-30}$ seconds after the big bang, when the energy of the Universe was $10^{12}$GeV. This candidate is till now one of the most accepted candidates for the nature of dark matter, if its abundance is about $10^9$ particles per cubic centimetre.

The first in suggesting that a dark halo could be
a Bose-Einstein condensate were \citet{sang-jin} and \citet{ji}
who used the weak field limit to show that a Bose-Einstein
Condensate (BEC) with several nodes can fit the rotation galaxy curves
with a very good accuracy. Further investigations on this
direction were performed by \citet{joe-weon}, where they
incorporated $\phi^4$ interactions to the scalar field potential
and used the Gross-Pitaevskii equation instead of the Schr\"odinger
one \citep{joe}. Nevertheless,
\citet{seidel,seidela} showed that when the whole BEC is in the ground state, many nodes in
Einstein-Klein-Gordon fields are unstable, since they evolve into
the 0-node solution after a while (for a clear explanation to this
point see also \citep{fcoluis}). Thus, the static solutions  given
by \citet{sang-jin, ji, joe-weon} are expected to be unstable.

Later on, \citet{peebles99} proposed that a scalar field driven by
inflation can behave as a perfect fluid and can have interesting
observational consequences in structure formation. Besides that, they
performed a sound waves analysis of this hypothesis giving
some qualitative ideas for the evolution of these fields and
called it fluid dark matter \citep{peea, peeb}.
Independently and in an opposite way,
\citet{matosguzman99} proposed a scalar field coming from
some unify theory can condensate and collapse to form haloes of
galaxies. Very early, this scalar field behaves as a perfect
fluid, however its ultralight mass causes that the bosons condensate
at very high temperature and collapse in a very different way as
the fluid dark matter of \citet{peebles99} did. They were able to fit
reasonably rotation curves of some galaxies using an exact solution of the
Einstein equations with an exponential potential
\citep{matosguzman99, gusmanmatos00, bernal}. The first
cosmological study of the SFDM was performed in
\citet{matosurena00a, matosurena00b} where a $cosh$ scalar
field potential was used. The cosmology reproduces all features of
the $\Lambda$ Cold Dark Matter ($\Lambda$CDM) model in the linear regime of
perturbations.

On the other hand, \citet{julien} and \citet{arbey} used a
complex scalar field with a quartic potential
$m^2\phi\phi^{\dag}+\lambda(\phi\phi^{\dag})^2$ and solved
perturbations equations (weak field limit approximation) to fit the
rotational curves of dwarf galaxies with a very good accuracy,
provided that $m^4/\lambda\sim50-75$ eV$^4$. 

The importance of scalar fields in the dark sector has been increased, for instance,
several authors have investigated the unification
of dark matter and dark energy in a single scalar field
\citep{pad,arbey,bert}. Recently
\citet{liddleurena06,liddleurena08} proposed
that the landscape of superstring theory can provide the Universe with a $\phi^2+\Lambda$
scalar field potential. Such scalar field can inflate the Universe
during its early epoch, after that, the scalar field can decay into
dark matter. The constant $\Lambda$ can be interpreted as the
cosmological one. This model could explain all unknown components of
the Universe in a simple way. Another interesting
model in order to explain the scalar fields unification, 
dark sector and inflation, is using a complex scalar field 
protected by an internal symmetry \citep{perez}.

In the present work the main idea is that if scalar fields are
fundamental, they live as unified fields in some very early moment
at the origin of the Universe. As the Universe expands, the scalar
fields cool together with the rest of the particles until they
decouple from the rest of the matter. After that, only the
expansion of the Universe will keep cooling the scalar fields. If
the scalar field fluctuation is under the critical temperature of
condensation, the object will collapse as a BEC. After inflation,
primordial fluctuations cause that the scalar fields collapse and
form haloes of galaxies and galaxy clusters. The cooling of
scalar fields continue till the fluctuation separates from the
expansion of the Universe.

In this work we study the most simple model of SFDM, using a
$\phi^2$ scalar field potential. In sections \ref{sec:estadistica}
and \ref{sec:einstein} we review the statistic of a boson gas to
condensate and form a BEC, focusing in the necessary features for
the BEC to form a halo of a galaxy and integrate the Einstein
equations with a BEC matter. In section \ref{sec:cosmologia} we
transform the Einstein field equations into a dynamical system, then we
numerically integrate them and look for the atractor points. We give
some conditions on how these field equations can give the right
behavior to reproduce the Universe we observed. Finally, in section
\ref{sec:conclusiones} we conclude that this SFDM model could
explain the dark matter of the Universe.

\section{The statistic of a BEC}\label{sec:estadistica}

In this section we review the condensation of an ideal Bose gas
of $N$ particles with mass $m$ contained
in a volume $V$ with temperature $T$ and with only a portion $\rho_0$ of the system in the ground state. 
In order to see that and to be self contained, let us start from its
grand partition function $\cal Q$, which is given by
\begin{equation}
{\cal Q}(z,V,T)=\prod_{\bf p}\frac{1}{1-ze^{-\beta\epsilon_{\bf
p}}},
\end{equation}
where the fugacity $z \equiv e^{\beta\mu}$ is defined in terms of the 
chemical potential $\mu$ and $\beta \equiv 1/T$. In this paper we use
the fundamental constants $\hbar=c=k_B=1$. 

Then, the equation of state for an ideal Bose gas is
\begin{equation}
\frac{PV}{T}=\log{\cal Q}=-\sum_{\bf
p}\log(1-ze^{-\beta\epsilon_{\bf p}})\label{PV}.
\end{equation}
\\
Thus, the grand partition function directly gives the pressure P as a 
function of $z$, $V$, and $T$.
\\
On the other hand the particle number $N$ and the internal energy $U$ are
\begin{eqnarray}
N&=&z\frac{\partial}{\partial z}\log{\cal Q}=\sum_{\bf
p}\frac{ze^{-\beta\epsilon_{\bf p}}}{1-ze^{-\beta\epsilon_{\bf
p}}},\\
U&=&-\frac{\partial}{\partial \beta}\log{\cal Q}=\sum_{\bf
p}\frac{\epsilon_{\bf p}ze^{-\beta\epsilon_{\bf
p}}}{1-ze^{-\beta\epsilon_{\bf p}}},\label{U}
\end{eqnarray}
where  $\epsilon_{\bf p}$ is the single-particle energy with momentum ${\bf p}$
and the average occupation
numbers $<n_{\bf p}>$ are given by
\begin{equation}
<n_{\bf p}>=\frac{ze^{-\beta\epsilon_{\bf
p}}}{1-ze^{-\beta\epsilon_{\bf p}}}, \label{np}
\end{equation}
\\
which satisfy the conditions
\begin{eqnarray}
N&=&\sum_{\bf p}<n_{\bf p}>,\label{np1}\\
U&=&\sum_{\bf p}\epsilon_{\bf p}<n_{\bf p}>.\label{np1}
\end{eqnarray}
 
Now we let $V\to 0$ taking the limit of continuity, 
and replace sums over {\bf p} by integrals
over {\bf p}, then we obtain the following equation of state
\begin{eqnarray}
\frac{PV}{T}&=&-\frac{2 V}{(2\pi)^2}\int_0^{\infty}dp\ p^2\log(1-ze^{-\beta p^2/2m})-\log(1-z), \nonumber \\
N&=&\frac{2 V}{(2\pi)^2}\int_0^{\infty}dp\ p^2\frac{ze^{-\beta p^2/2m}}
{1-ze^{-\beta p^2/2m}}+\frac{z}{1-z}.\\
\nonumber
\end{eqnarray}
\\
These equations can be written into the equivalent form

\begin{eqnarray}
\frac{PV}{T}&=&\frac{V}{\lambda^3}g_{5/2}(z)-\log(1-z),\label{eq2}\\
N&=&\frac{V}{\lambda^3}g_{3/2}(z)+\frac{z}{1-z},\label{N2}\\
\nonumber
\end{eqnarray}
where $\lambda=\sqrt{2\pi/mT}$ is the thermal wavelength,
and
\begin{eqnarray}
 g_{5/2}(z)&=&-\frac{4}{\sqrt{\pi}}\int_0^{\infty}dx\
x^2\log(1-ze^{-\beta x^2}),\nonumber\\
g_{3/2}(z)&=&z\frac{\partial}{\partial z}g_{5/2}(z) \label{gs}.
\end{eqnarray}

Moreover, the internal energy is found from the formulas
(\ref{PV}) and (\ref{U})
\begin{equation}
U=\frac{3}{2}\frac{TV}{\lambda^3}g_{5/2}(z),\label{U_fin}
\end{equation}
and as consequence the relation U=3/2PV is fulfilled.
\\
From equation (\ref{np}) we see that
\begin{equation}
<n_{0}>=\frac{z}{1-z}, \label{n0}
\end{equation}
which is the average occupation number for a single particle with
occupation level ${\bf p}=0$. Equation (\ref{N2}) can also be
written as
\begin{equation}
\lambda^3\frac{<n_{0}>}{V}=\lambda^3\frac{N}{V}-g_{3/2}(z).\label{N3}
\end{equation}
\\
This equation tell us that $\frac{<n_{0}>}{V}>0$ and therefore
the temperature and the specific volume are such that
$\lambda^3\frac{N}{V}>g_{3/2}(z)$. This means that a finite fraction of the 
particles will be in the ground state with ${\bf p}=0$, $i.e.$,
the Bose gas condensates. 
In the region of condensation, the fugacity
$z\sim1$ and the function $g(z)$ goes to the Riemann $\zeta$
function $g_{l}(z)\rightharpoonup\zeta(l)$.

The thermodynamical surface which
separates the condensation region from the rest of the $P-V-T$ space, is given by
\begin{equation}
\lambda_c^3\frac{N}{V}=g_{3/2}(1)=2.612,\label{N4}
\end{equation}
thus $\lambda_c$ can be interpreted as the value for which the
thermal wavelength is of the same order of magnitude as the
average interparticle separation. Equation (\ref{N4}) defines the
critical temperature for which the Bose Condensate forms. This
temperature is given by
\begin{equation}
T_c=\frac{2\pi}{m_{\phi}^{5/3}}\left(\frac{\rho}{g_{3/2}(1)}\right)^{\frac{2}{3}},\label{T_crit}
\end{equation}
where $\rho={m_{\phi}N}/{V}$ is the density of the Bose gas. At
constant temperature, equation (\ref{T_crit}) defines a critical
density
\begin{equation}
\rho_c=\frac{m_{\phi}g_{3/2}(z)}{\lambda^3}\label{rho_crit}.
\end{equation}
\\
Thus, the region of condensation of the Boson gas is determined by
$T<T_c$ or $\rho>\rho_c$.

After the Bose gas condensates most of the bosons lie in the
ground state, the scalar field starts to oscillate around the
minimal of its potential and the scalar field starts to behave as
dust \citep{turner}. Thus, after the scalar field decouples from
the rest of the matter, the temperature of the BEC goes like

\begin{equation}
T_{BEC}=T^{(0)}_{BEC}\left(\frac{a_{0}}{a}\right)^2\label{T_BEC},
\end{equation}
where $T^{(0)}_{BEC}$ is the actual temperature of the BEC,
$a$ is the scale factor of the Universe and $a_{0}=1$ is
the value of the scale factor at present.
\\
In the same way, as the BEC behaves as matter, its density goes
like $\rho_{BEC}=\rho^{(0)}_{BEC}/a^3$, where $\rho^{(0)}_{BEC}$ is the
actual matter content of BEC in the Universe. With this result,
equation (\ref{T_crit}) can also be transformed into

\begin{eqnarray}
T_{c}&=&\frac{2\pi}{\,m_{\phi}^{5/3}}
\left(\frac{\Omega^{(0)}_{BEC}\rho_{crit}}{\zeta(3/2)}\right)^{\frac{2}{3}}\frac{1}{a^2},\\
&=&
6.2\times10^{-31}\frac{(\Omega^{(0)}_{BEC}h^2)^{2/3}}{(m_{\phi}/\textrm{GeV})^{5/3}}\frac{1}{a^2}\textrm{GeV}
\label{T_BEC_Num},
\end{eqnarray}
where $\Omega^{(0)}_{BEC}$ is the actual rate of BEC,
$\rho_{crit}$ is the critical density of the Universe,
$h \equiv H_{0}/(100 \textrm{km\,s}^{-1}\textrm{Mpc}^{-1})$ being $H_{0}$
the actual value of Hubble's parameter.

If the actual standard model of particles could be extended
to higher temperatures, we have to expect that the scalar field
which forms the BEC, interacts with the rest of the particles to a
temperature over some temperature $T_s$. Because the physics of
elemental particles is well known till temperatures like GeV, we do
not expect that an exotic particles as these scalar fields appear
under temperatures like TeV. Here we have two possibilities, the first one is that the scalar field has never had interaction with the rest of the particles and it evolves independently from the rest of the fields, with only a gravitational interaction. In this case the scalar field condensates at the beginning of the Universe. The second possibility is that in the early universe the scalar field lived unified with the rest of the particles in a thermal bath and at some moment of its evolution, separates from the interaction. If this is the case let us suppose here that the scalar
field which forms the BEC decouples from the rest of the matter at a
temperature over TeV. Under this temperature, this scalar field has
almost no interaction with the rest of the matter. If we expect that
this scalar field forms a BEC, its critical temperature must be
lower than the temperature of the scalar field decoupling. This fact
gives us an upper bound of the mass $m_{\phi}$ of the scalar field
\begin{equation}
m_{\phi}<10^{-17}\textrm{eV}\label{mphi_bound}.
\end{equation}

On the other hand, from numerical simulations \citep{seidel}
we know that scalar fields form gravitationally bounded
objects with a critical mass given by
\begin{equation}
M_{crit}\sim \tilde{m} \frac{m_{pl}^2}{m_{\phi}}\label{Mcrit},
\end{equation}
where $m_{pl}$ is the Planck mass and $\tilde{m}$ is a factor such that
$\tilde{m}\approx0.6$ for both complex scalar fields (boson stars) and
real scalar fields (oscillatons). With the value given in
(\ref{mphi_bound}), the scalar field can form a gravitationally
bounded BEC with a critical mass given by
\begin{eqnarray}
M_{crit}&>&1.491\times10^{64}\textrm{GeV},\\
&=&2.658\times10^{40}\textrm{gr},\\
&=&13.36\times10^{6}M_{\odot}.\label{Mcrit_Num}
\end{eqnarray}

This is an interesting result, if there exists a scalar field and
plays any role in the Universe at this moment, this scalar field
must have a mass lower than the mass given in (\ref{mphi_bound})
and they are forming gravitationally bounded BECs with masses
around the mass given in \citet{colapso}.

\section{Self-gravitating BEC}\label{sec:einstein}

In this section we give some general features of the gravitational
collapse of the BEC, we only pretend to show a generic behavior of
any self-gravitating BEC. The BEC cosmology have been studied by
\citet{fukuyama} and many numerical simulations of this
collapse are given in \citet{colapso,colapsob,apjpaco}
and besides. \citet{fcoluis}
found that a BEC in the ground state  are very
stable under different initial conditions. After
the Bose gas condensates the gravitational force makes the gas
collapse and form self-gravitating objects. Let us suppose that
the halo is spherically symmetric, which could not be to far from
the reality. In that case, the space-time metric reads
\begin{equation}
ds^2=-e^{2\nu}dt^2+\frac{dr^2}{1-\frac{2MG}{r}}+r^2d\Omega^2\label{metric},
\end{equation}
where the function $\nu=\nu(r)$ is essentially the Newtonian
potential and $M=M(r)$ is the mass function given by
\begin{eqnarray}
M&=&4\pi\int\rho\ r^2\,dr, \nonumber\\ 
\frac{d\nu}{dr}&=&G\frac{M+4\pi r^3\
P}{r^2\left(1-\frac{2MG}{r}\right)}\label{nu}.
\end{eqnarray}

The Einstein field equations reduce to equations (\ref{nu}) and the
Oppenheimer-Volkov equation
\begin{equation}
\frac{dP}{dr}=-G\frac{(P+\rho)(M+4\pi r^3\
P)}{r^2(1-\frac{2MG}{r})}\label{OppVol}.
\end{equation}
\\
Let us focus in the case when the gas is far from forming a black
hole. In that case we suppose that $2MG<<r$ and equation
(\ref{OppVol}) reduces to
\begin{equation}
\frac{dP}{dr}=-4\pi G\,r\ P(P+\rho)\label{OppVolRe}.
\end{equation}
\\
The equation of state can be obtained from the equation $PV=2/3\,U$,
(\ref{N2}) and (\ref{U_fin}). Combining all equations we obtain
that
\begin{eqnarray}
P&=&\frac{2\pi}{m_{\phi}^{8/3}}\frac{g_{5/2}(z)}{g_{3/2}(z)^{5/3}}(\rho-\rho_0)^{5/3},\\
&=&\omega(\rho-\rho_0)^{5/3},\label{eq_est}
\end{eqnarray}
where $\omega$ is the constant
\begin{equation}
\omega \equiv \frac{2\pi}{m_{\phi}^{8/3}}\frac{g_{5/2}(z)}{g_{3/2}(z)^{5/3}}\label{omega},
\end{equation}
and $\rho_0=m_{\phi}<n_0>/V$ is the mean density of the particles
in the ground state. Thus, the
Oppenheimer-Volkov equation (\ref{OppVol}) transforms into
\begin{equation}
\frac{d\rho}{dr}=-\frac{12}{5}\pi G r
(\rho-\rho_0)(\omega(\rho-\rho_0)^{5/3}+\rho)\label{OppVolRe3}.
\end{equation}

\begin{figure}
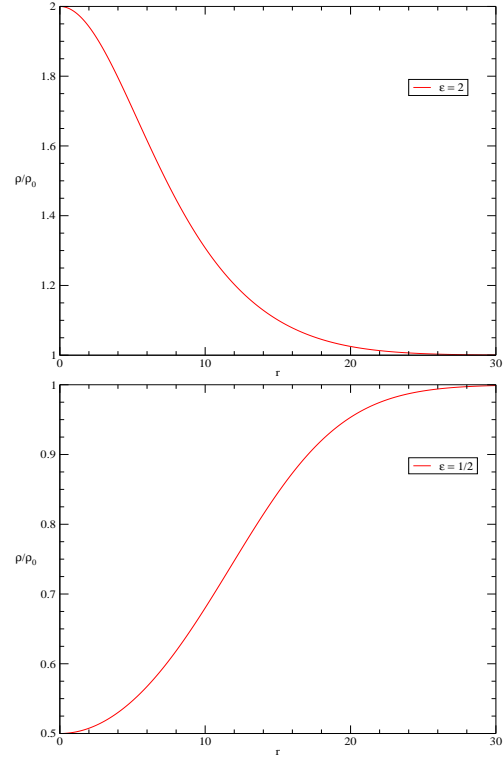

\centerline{\epsfysize=5cm \epsfxsize=6.5cm \epsfbox{r2.eps}}
\centerline{\epsfysize=5cm \epsfxsize=6.5cm \epsfbox{r1_2.eps}} \caption{Plot of
the $\rho(r)$ function given in equation (\ref{sol}) for $\epsilon<1$
(top plot) and for $\epsilon>1$ (down plot). The plot is done in terms
of $\rho(r)/\rho_0$. We have set $\epsilon=2$ and $\epsilon=1/2$ for each
plot, respectively and $\rho_0=0.002$.} \label{fig:BEC1}
\end{figure}

This differential equation can be easily numerically solved.
Nevertheless, we have two interesting limits of equation
(\ref{OppVolRe3}). First suppose that the $\omega$ constant is
small such that $P<<\rho$. This situation occurs for big scalar
field masses $m_{\phi} \sim m_{Planck} $. In that case, the
equation (\ref{OppVolRe3}) contains an analytical solution given
by
\begin{equation}
\rho(r)=\frac{\rho_0}{1-\left( 1- \frac {\rho_0}{\rho(0)}
\right)e^{-\frac{6}{5}\pi G \rho_0r^2} }\label{sol},
\end{equation}
where $\rho(0)$ is the central density of the BEC.
Observe that when $r\rightharpoonup\infty$, the function
$\rho(r)\rightharpoonup\rho_0$. For numerical convenience we set
$\rho(0) = \epsilon \rho_{0}$ in the plot, being $\epsilon$
a constant. The function changes dramatically
for different values of $\epsilon$. If $\epsilon>1$, the density $\rho(r)$
decreases, but if $\epsilon<1$ the density increases. The behavior of
the density is shown in Fig. \ref{fig:BEC1}. 
This means that if the central density of the BEC is bigger than
the density of the ground state, we have the upper profile in Fig.
\ref{fig:BEC1},  but if it is less than it, we have the bottom
profile.

\begin{figure}
\centerline{\epsfysize=6.5cm \epsfbox{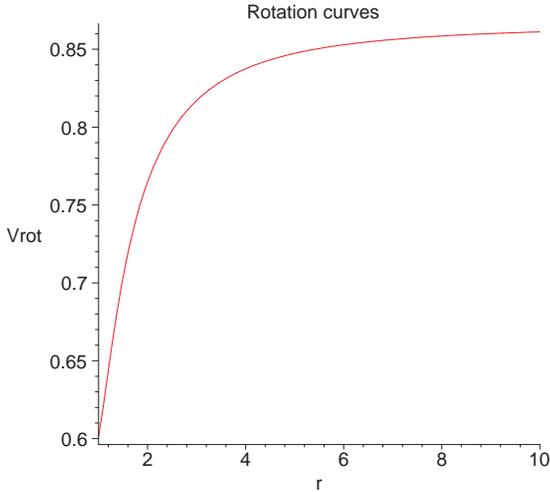}} \caption{
Rotation curve derived from metric
(\ref{sol_metric}). The velocity and the coordinate $r$ are in
arbitrary units.}\label{fig:Rot_cuv}
\end{figure}

The second and for us, a more interesting limit
of equation (\ref{OppVolRe3}) is when $P>>\rho$. This occurs when
the scalar field mass is small enough $m_{\phi}<<m_{Planck}$, as
for astrophysical BEC. In this limit the Oppenheimer-Volkov
equation has also an analytical solution given by
\begin{eqnarray}
\rho(r)&=&\frac{\rho(0)-\rho_0}{(2\pi G
r^2\omega(\rho(0)-\rho_0)^{5/3}+1)^{3/5}}+\rho_0,\nonumber\\
&=&\left(\frac{p(0)/\omega}{2\pi G r^2p(0)+1}\right)^{3/5}+\rho_0,
\label{sol_Astroph}
\end{eqnarray}
or equivalently $P=1/(2\pi G r^2+1/P(0))$. In this case the pressure
dominates the BEC, the pressure acquire a maximum for $P(0)$.
Far away enough from the center of the BEC we can approximate
equation (\ref{sol_Astroph}) with
\begin{equation}
\rho=\left(\frac{1/\omega}{2\pi G r^2}\right)^{3/5}+\rho_0
\label{sol_Astroph1},
\end{equation}
which implies a space-time metric for the BEC given by
\begin{equation}
ds^2=\frac{dr^2}{1-2(r_0r^{4/5}+\frac{4}{3}\pi G \rho_0r^2)}-exp(2\nu)dt^2+r^2d\Omega^2
\label{sol_metric},
\end{equation}
where $r_0 \equiv 10/9(4\pi^2/\omega^3)^{1/5}$. Function $\nu$ determines
the circular velocity (the rotation curves) $V_{rot}$ of test
particles around the BEC. Using the geodesic equation of metric
(\ref{sol_metric})  one obtains that $V_{rot}^2=rg_{tt,r}/(2g_{tt})=r\,\nu'$
\citep{todo}. Using equations (\ref{nu}) we can integrate
the function $\nu$ and obtain the rotation curves. The plot is
shown in Fig. \ref{fig:Rot_cuv}, where we see that the form 
of the rotation curves are analogous as the expected from the
observed in galaxies, specially in LSB and dwarf ones
\citep{blok1,blok2,simon} besides SFDM predicts 
a core density profile that could have some astrophysics advantages \citep{javier} 
over the standard model (cuspy profiles). However, the 
discussion of the central region of the rotation curves continue.
This is the main reason why it is not convenient to try self-gravitating BECs
in the Newtonian limit. Remain that the Newton theory can be
derived from the Einstein one for slow velocities, weak fields and
pressures much smaller than the densities. However these last conditions
is not fulfilled in self-gravitating BEC.

From these results and from the simulations given in
\citet{fcoluis} it follows a novel paradigm for structure
formation, which is different from the bottom-up one. In the SFDM
paradigm, after the big bang the scalar field expands till
decouples from the rest of the matter. If the scalar field has
sufficient small mass such that its critical temperature of
condensation is less than the temperature of decoupling, the
scalar field forms a BEC. Then the scalar field collapses forming
objects which final mass is not bigger than the critical mass
$m_{Planck}^2/m_{\phi}$. These objects contain a density profile very
similar to the profile shown in the top of Fig. \ref{fig:BEC1}.
They are very stable under perturbations. It has been proposed
that the dark matter in galaxies and clusters is a scalar field
with a mass of $10^{-22}$eV \citep{colapso}. If this
were the case, the main difference for the structure formation of
this ultralight scalar field with the bottom-up paradigm is that
the SFDM objects form just after the collapse of the scalar field
and remain so during the rest of the Universe expansion.
Furthermore, they can collide together but after the collision the
objects remain unaltered, since they behave like solitons \citep{headon}.
This means that in a merging of BEC they pass through each other
without some alterations in its total mass as collisionless dark matter.
This paradigm implies then that we must be able to see well formed
galaxies with the actual masses for very large redshifts, longer
than those predicted by the bottom-up paradigm, $i.e.$, by CDM. 
In this sense some authors \citep{cimatti} suggest a discrepancy between the observed 
population of massive spheroidal galaxies at high redshift with the 
numerical simulations of hierarchical merging in a $\Lambda$CDM 
scenario that underpredict this population. However, 
the discussion continues because other physical processes, 
as feedback, could have important effects in this galaxies.

\section{The Cosmology}\label{sec:cosmologia}

In this section we review the Cosmology given by a SFDM model with
two different scalar field potentials: $V(\phi)=\frac{1}{2}\,m^2\phi^2$
and $V(\phi)=V_0\left[\cosh({\kappa}\lambda\phi)-1\right]$
where $m$ is the mass of the boson particle,
$V_0$ and $\lambda$ are free parameters fixed
with cosmological data and $\kappa^2=8\pi G$. Based on the current observations
of 5-year WMAP data \citep{hin} we will consider a Universe
evolving in a spatially-flat Friedmann Lema\^itre-Robertson-Walker
spacetime. We assume that this Universe contains a real scalar field
($\phi$) as dark matter, radiation (r), neutrinos ($\nu$),
baryons (b) and a cosmological constant ($\Lambda$) as dark energy.
\\
The total energy density of a homogeneous scalar field is given by
\\
\[\rho_{\phi}=\frac{1 }{2}\dot\phi ^2+V(\phi), \]
the radiation and baryonic components are represented by
perfect fluids with baryotropic equation of state
$p_{\gamma}=(\gamma-1)\rho_{\gamma}$, where $\gamma$ is a constant,
$0\le \gamma \le 2$. For example, for radiation and neutrinos
($\gamma_{r,\nu}=\frac{4}{3}$), for baryons ($\gamma_{b}=1$) and
finally for a cosmological constant ($\gamma_{\Lambda}=0$).

\vspace{0.5cm}
Thus, the field equations for a Universe with these
components are given by
\begin{eqnarray}
\dot H&=&-\frac{\kappa^2}{2}(\dot\phi^2+\gamma \rho_{\gamma}),\nonumber\\
{\ddot \phi} &+& 3\,H {\dot \phi} + \partial_{\phi}V=0,\nonumber\\
{\dot\rho_{\gamma}}&+&3\,\gamma\,H\,\rho_{\gamma}=0, \label{eq:ecua}
\end{eqnarray}
and the Friedmann equation

\begin{equation}
H^2=\frac{\kappa^2}{3}\left(\rho_{\gamma}+\frac{1}{2}\dot \phi^2
+V(\phi) \right)\label{eq:Frie}.
\end{equation}

In order to analyze the behavior of the different components of this
Universe, we are going to use the dynamical system formalism 
following Appendix \ref{App:DynSys}.

\subsection{The $\phi^{2}$ scalar potential}

We start our cosmological analysis of SFDM taking the potential
\begin{equation}
V(\phi)=\frac{1}{2}\,m^2\phi^2 \label{eq:phi2},
\end{equation}
and developing the standard procedure to transform it into a
dynamical system. For doing so, the new variables (\ref{eq:varG})
for the system of equations (\ref{eq:ecua}) read
\begin{eqnarray}
x&\equiv& \frac{\kappa}{\sqrt{6}}\frac{\dot\phi}{H},\,\,\,
u\equiv \frac{\kappa}{\sqrt{6}}\frac{m\phi}{H},\nonumber\\
z_{\gamma}&\equiv& \frac{\kappa}{\sqrt{3}}\frac{\sqrt{\rho_{\gamma}}}{H}.\label{eq:var}
\end{eqnarray}

Using the definitions given in (\ref{eq:var}), the evolution
equations (\ref{eq:ecua}) for potential (\ref{eq:phi2}) transform
into an autonomous system

\begin{eqnarray}
x'&=& -3\,x - \frac{m}{H} u+\frac{3}{2}\Pi\,x,
\nonumber\\
u'&=& \frac{m}{H}x + \frac{3}{2}\Pi\,u,
\nonumber\\
z_{\gamma}'&=&\frac{3}{2}\left(\Pi-\gamma \right)\,z_{\gamma},
\nonumber\\
-\frac{\dot H}{H^2}&=&\frac{3}{2}(2x^2+\gamma z_{\gamma}^2)\equiv
\frac{3}{2}\Pi  \label{eq:sd},
\end{eqnarray}
\\
where as in Appendix \ref{App:DynSys}, prime denotes a derivative
with respect to the e-folding number $N=\ln(a)$. Again the choice of
phase-space variables (\ref{eq:var}) transforms the Friedmann
equation into a constraint equation
\begin{equation}\label{eq:fri}
F\equiv x^2+u^2+z_{\gamma}^2=1.
\end{equation}

Because we are considering an expanding Universe which implies that
$H>0$ and from the variable definitions (\ref{eq:var}), we can see
that $u,z_{\gamma} \geq 0$. With these variables, the density
parameters can be written as
\begin{eqnarray}\label{eq:dens}
\Omega_{DM}&=&x^2+u^2, \nonumber\\
\Omega_{\gamma}&=&z_{\gamma}^2,\nonumber\\
\Omega_{\Lambda}&=&l^2,
\end{eqnarray}
where we have added explicity a cosmological constant variable $l
\equiv z_{\Lambda}$. Moreover, with the physical constraint
$0\le \Omega \le 1$ and the Friedmann equation
$\Omega_{DM}+\Omega_{\gamma}+\Omega_{\Lambda}=1$ the variable space
is bounded by
\[
0\le x^2+u^2+z_{\gamma}^2+l^2\le 1.
\]

On the other hand, observe that the variable space (\ref{eq:sd}) is
not a completely autonomous one because $H$ is an external
parameter. In order to close the system we define a new variable $s$
given by
\begin{equation}\label{eq:ssss}
s\,\, \equiv \,\,\frac{m}{H},
\end{equation}
which dynamical equation (\ref{eq:sdG_H}) is
\begin{equation}
s'=\frac{3}{2}\Pi\,s \nonumber.
\end{equation}
\\
With this new variable, system (\ref{eq:sd}) is now an autonomous
one. The whole close system is
\begin{subequations}
\begin{eqnarray}
x'&=& -3\,x - s u+\frac{3}{2}\Pi\,x \label{eq:sdp2_x},\\
u'&=&s\,x + \frac{3}{2}\Pi\,u  \label{eq:sdp2_u},\\
z_{\gamma}'&=&\frac{3}{2}\left(\Pi-\gamma\right)\,z_{\gamma} \label{eq:sdp2_z},\\
l'&=&\frac{3}{2}\Pi\,l, \\
s'&=&\frac{3}{2}\Pi\,s \label{eq:sdp2_s}.
\end{eqnarray}\label{eq:sdp2}
\end{subequations}

In order to acquire geometrical information that dynamical system
analysis provide (see Appendix \ref{App:DynSys}), we study the
stability of (\ref{eq:sdp2}). To do this, we define the vector
$\vec{x}=(x,u,z_{\gamma},l,s)$ and consider a linear perturbation of
the form $\vec{x}\to \vec{x_{c}}+\delta \vec{x}$. The linearized system
reduces to $\delta \vec{x'} =\mathcal{M} \delta \vec{x}$, where
$\mathcal{M}$ is the \textit{Jacobian matrix} of $\vec{x'}$ and it
reads as

\vspace{0.5cm}

\[\mathcal{M} =
{\footnotesize \left(
\begin{array}{ccccc}
\frac{3}{2}\Pi-3+6x^2 & -s & 3 \gamma x\, z & 0 & -u \\
6 x\,u + s& \frac{3}{2}\Pi & 3 \gamma \,u\,z & 0 & x\\
6 x\,z & 0 & \frac{3}{2}\Pi+ 3 \gamma z^2-\gamma & 0 &  0\\
6 x\,l & 0 & 3 \gamma\,\,l\,\, z & \frac{3}{2}\Pi &  0\\
6 x\,s & 0 & 3 \gamma \,s\,z & 0 & \frac{3}{2}\Pi\\
\end{array} \right).}
\]
\vspace{0.5cm}

The equilibrium points $\vec{x_c}$ of the phase space
$\{x,u,z_{\gamma},l,s\}$, considering only $\gamma = 4/3$, are
then
\begin{enumerate}
\item $\{\pm 1, 0,0,0,0\}$ Kinetic scalar domination
\item $\{0,0,1,0,0\}$ Radiation domination
\item $\{0,0,0,1,s\}$ Cosmological constant domination
\item $\{0,u,0,l,0\}$ Cosmological constant and Potential scalar
domination
\end{enumerate}
Finally, the eigenvalues of the matrix $\mathcal{M}$ valued at
the critical points listed above read
\begin{enumerate}
\item $\{6,3,3,3,3-\gamma \}$
\item $\{\frac{3\gamma}{2},\frac{3\gamma}{2},\frac{3\gamma}{2},\frac{7\gamma}{2},
\frac{3}{2}(-2+\gamma)\}$
\item $\{0,0,\frac{1}{2}(-3-\sqrt{9-4s^2}),\frac{1}{2}(-3+\sqrt{9-4s^2}),-\gamma \}$
\item $\{-3,0,0,0,-\gamma \}$
\end{enumerate}

\begin{figure}
\centerline{\epsfysize=5cm \epsfxsize=6.5cm \epsfbox{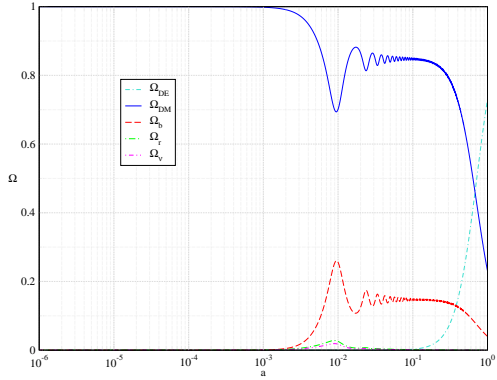}} \caption{Evolution
of the density parameters for the system of equations (\ref{eq:sdp2}).
The plot shows the dark matter domination epoch at early times, a source point.
The cosmological constant at the future of the Universe is an attractor
point.} \label{fig:normal}
\end{figure}

As we can see, the radiation domination epoch shows a saddle
point, however, in order to reproduce the big bang nucleosynthesis
process is necessary that this kind of matter would had dominated
in the past of the Universe. In other words, the radiation points
should have corresponded to a source point. The domination of dark
matter in the past (a source point) and the cosmological constant
in the future (an attractor point) are showed in the Fig.
\ref{fig:normal}.

In the following, we integrate system (\ref{eq:sdp2}) with the
constraint (\ref{eq:fri}), following the procedure shown in
Appendix \ref{App:DynSys}. In general this system is very difficult
to integrate because it is a non-linear four-dimensional
differential system of equations. It is clear that system
(\ref{eq:sdp2}) is a complete system which can fulfill or not the
constraint (\ref{eq:fri}). However, as it was shown in Appendix
\ref{App:DynSys} system (\ref{eq:sdp2}) together with constraint
(\ref{eq:fri}) is completely integrable.
For simplicity we will take all the perfect fluid components as the
equation $z_{\gamma}'=3/2(\Pi-\gamma)z_{\gamma}$ with the Friedmann
equation $x^2+u^2+z_{\gamma}^2=1$.
\\
Thus, we substitute $3/2\Pi$ from equation (\ref{eq:sdp2_s}) into
the rest of the equations. With this substitution equation
(\ref{eq:sdp2_z}) integrates in terms of $s$ as
\begin{equation}
z_{\gamma}=\sqrt{\Omega_{\gamma}^{(0)}}\,s\,\exp(-\frac{3}{2}\gamma\,N)\label{eq:lfinal},
\end{equation}
where $\Omega_{\gamma}^{(0)}$ is an integration constant. We multiply
(\ref{eq:sdp2_x}) by $2\,x$ and (\ref{eq:sdp2_u}) by $2\,u$ and sum
both equations. We obtain

\begin{equation}
(x^2+u^2)'=-6\,x^2 +2 \ln(s)'(x^2+u^2)\label{eq:x2+u2}.
\end{equation}
\\
Now, we use constraint (\ref{eq:fri}) and equation (\ref{eq:lfinal})
into equation (\ref{eq:x2+u2}) to obtain
\begin{equation}
6\,x^2=2\,\ln(s)'-3\gamma\,s^2\,\Omega_{\gamma}^{(0)}\exp(-3\,\gamma\,N)\label{eq:x2}.
\end{equation}

We substitute (\ref{eq:x2}) and (\ref{eq:lfinal}) into
(\ref{eq:sdp2_s}) to obtain $0=0$. Therefore, $s$ is not an independent
variable and we cast it into the system as a control variable
which parametrizes the decrease of H,
a similar result is found by \citet{mayra}.
In what follows we will use this important result.

Of course, to guess variable $s$ in order to fulfill constraint
(\ref{eq:fri}) is not so easy. In order to avoid this problem we can
consider the observed dynamic for $H$ and model it by the following
ansatz
\begin{equation}\label{eq:ants}
H\, \equiv\, \frac{{t_0}^{n-1}}{t^n},
\end{equation}
because it is well-know the behavior for $H$ at different epochs
\begin{equation}
H_{\textrm{dust}}=\frac{2}{3t},\,\,\, H_{\textrm{rad}}=\frac{1}{2t},\,\,\,
H_{\Lambda}=\sqrt{\frac{\Lambda}{3}}.
\end{equation}
\\
There exists a restriction in the parameter $n$. Because is well know that
$H$ is a function monotonically decreasing, $n$ has to satisfy $n\ge 0$.
With the ansatz (\ref{eq:ants}), the dynamical equation for $s$
reads
\begin{equation}\label{eq:ss}
s'\,\,=\,\,\left(m t_0\right)^{\frac{1}{n}-1} n\,\,
\left(\frac{1}{s}\right)^{\frac{1}{n}-2}=s_0\,s^{-k},
\end{equation}
where we have defined $k \equiv 1/n-2$.

In the following, we investigate if this system can reproduce the
observed Universe. We introduce the components of the background
Universe into the dynamical system described by (\ref{eq:sdp2})
adding to it baryons ($b$), radiation ($z$) and neutrinos ($\nu$).
Thus, the system transforms into
\begin{subequations}
\begin{eqnarray}
x'&=& -3\,x - s u+\frac{3}{2}\Pi\,x \label{eq:sd4_x},\\
u'&=& s x +\frac{3}{2}\Pi\,u \label{eq:sd4_u},\\
b'&=&\frac{3}{2}\left(\Pi -1\right)\,b \label{eq:sd4_b},\\
z'&=&\frac{3}{2}\left(\Pi-\frac{4}{3} \right)\,z \label{eq:sd4_z},\\
\nu'&=&\frac{3}{2}\left(\Pi-\frac{4}{3} \right)\,\nu \label{eq:sd4_n},\\
l'&=&\frac{3}{2}\Pi\,l \label{eq:sd4_l},\\
s'&=&s_0 \,s^{-k} \label{eq:sd4_s},
\end{eqnarray}\label{eq:sd4}
\end{subequations}
with $\Pi=2x^2+b^2+\frac{4}{3} z^2+\frac{4}{3} \nu^2$ and the
Friedmann equation reduces to the constraint
\begin{equation}\label{eq:fri4}
F=x^2+u^2+b^2+z^2+\nu^2+l^2=1.
\end{equation}
Using this ansatz we can reduce till quadratures the solution of
system (\ref{eq:sd4}). In order to do this, observe that
\[
\frac{3}{2}\Pi=s_0\,s^{-k-1}.
\]
Now, using this last identity, equation (\ref{eq:sd4_b})-
(\ref{eq:sd4_l}) can be integrated to give
\[
z_{\gamma}=z_0\,[s_0\,(k+1)\,N+s_1]^{\frac{1}{k+1}}e^{-\frac{3}{2}\gamma\,N},
\]
for each corresponding value of $\gamma$. Finally, equations
(\ref{eq:sd4_x}) and (\ref{eq:sd4_u}) can be integrated as
follows. We divide (\ref{eq:sd4_x}) by $x$ and (\ref{eq:sd4_u}) by
$u$ and take the difference between both equations. We define $y=x/u$
to obtain
\begin{equation}
y'+3y+q(N)y^2=-q(N)\label{eq:y},
\end{equation}
where function $q(N)=[s_0\,(k+1)\,N+s_1]^{{1}/{(k+1)}}$. Equation
(\ref{eq:y}) is a Riccati equation which can be reduce to a
Bernoulli equation by defining $y=w+y_1$, where $y_1$ is a known
solution of (\ref{eq:y}). It reduces to
\begin{equation}
w'+(3+2\,q\,y_1)\,w+q\,z^2=0\label{eq:z}.
\end{equation}
Equation (\ref{eq:z}) can be further reduced by defining $W=1/w$, we
obtain
\begin{equation}
W'-(3+2\,q\,y_1)\,W-q=0\label{eq:Z},
\end{equation}
which integral is
\begin{equation}
W = e^{A}\int{e^{-A}\,q\,dN} \label{eq:Zfin},
\end{equation}
with $A = \int{(3+2\,q\,y_1)\,dN}$. Thus
\begin{subequations}
\begin{eqnarray}
u &=& u_0\,q \exp{\left( \int{y\,q\,dN}\right)} \label{eq:sol_u},\\
x &=& x_0\,q \,e^{-3\,N} \, \exp{\left(-\int{\frac{q}{y}\,dN}\right)}   \label{eq:sol_x},\\
z_{\gamma}&=&z_0\,q\,e^{-\frac{3}{2}\gamma\,N}
\label{eq:sol_l},\\
y&=& \frac{1}{W}+y_1 \label{eq:sol_y}.
\end{eqnarray}\label{eq:sol}
\end{subequations}

In the particular case where $s_0=0$, the integrals can be solved
analytically, however this value for $s_0$ does not have a physical
meaning.

On the other hand, we can evaluate the integrals using
numerical methods for different values of the free constants. We can
obtain a numerical solution for the system using (\ref{eq:sol}) or
directly integrating system (\ref{eq:sd4}) with an
Adams-Bashforth-Moulton (ABM) method and using as initial data the
WMAP+BAO+SN recommended values to $\Omega^{(0)}_{\Lambda}=0.721$,
$\Omega^{(0)}_{DM}=0.233$, $\Omega^{(0)}_b=0.0454$,
$\Omega^{(0)}_r=0.0004$, $\Omega^{(0)}_{\nu}=0.0002$, the result is the same.
\begin{figure}
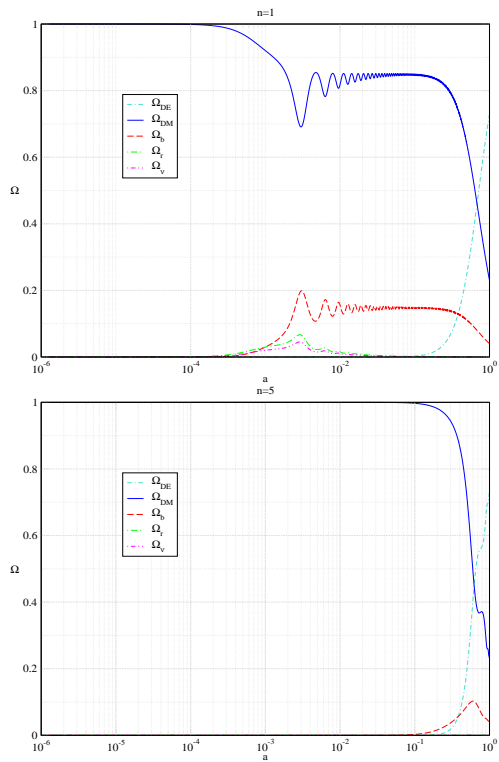

\centerline{\epsfysize=5cm \epsfxsize=6.5cm \epsfbox{n1.eps}}
\centerline{\epsfysize=5cm \epsfxsize=6.5cm \epsfbox{n5.eps}} \caption{Evolution
of the density parameters for the system (\ref{eq:sd4}) with
$n=1$ (top panel) and $n=5$ (bottom panel). This values of n are not reproduce
the standard behavior of $\Lambda$CDM} \label{fig:n1}
\end{figure}
\begin{figure}
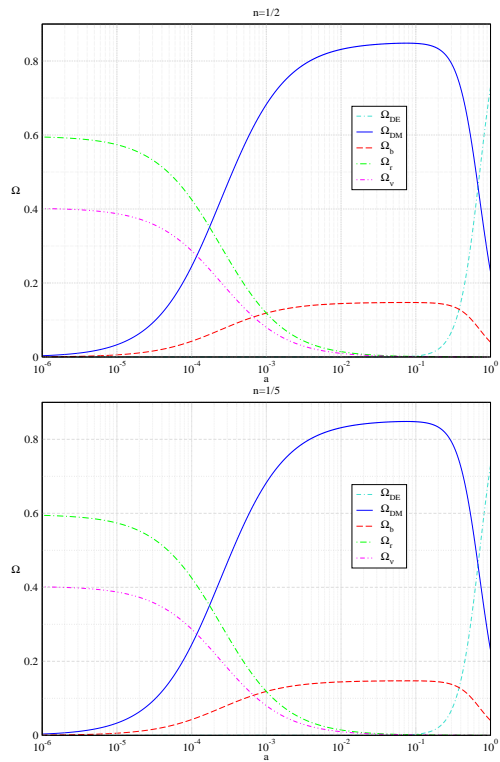

\centerline{\epsfysize=5cm \epsfxsize=6.5cm \epsfbox{n1_2.eps}}
\centerline{\epsfysize=5cm \epsfxsize=6.5cm \epsfbox{n1_5.eps}} \caption{ Upper panel:
evolution of the density parameters for the system (\ref{eq:sd4}) with $n=1/2$.
Lower panel: evolution of the density parameters for the system (\ref{eq:sd4})
with $n=1/5$. SFDM reproduces the standard $\Lambda$CDM behavior in both cases}
\label{fig:n2}
\end{figure}
\begin{figure}
\centerline{\epsfysize=5cm \epsfxsize=6.5cm \epsfbox{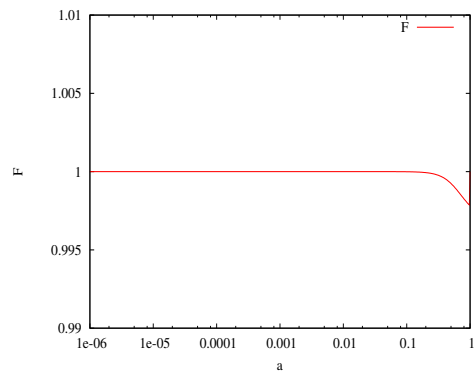}}
 \caption{Evolution of the function $F=x^2+u^2+b^2+z^2+\nu^2+l^2$ in
(\ref{eq:fri4}) for the system (\ref{eq:sd4}) with $n=1$, $5$,
$1/2$ and $1/5$. Function $F$ is exactly the same for all values
of $n$ in all these cases.} \label{fig:Pe}
\end{figure}

Fig. \ref{fig:n1} and Fig. \ref{fig:n2} show the numerical solutions
of the dynamical system (\ref{eq:sd4}). In Fig. \ref{fig:n1} we set
$n\geq1$, as examples we show $n=1,\,5$. From these figures it is
clear that the radiation remains subdominant for this values of $n$.
On the other hand, in Fig. \ref{fig:n2}, where the plots were made for
$n=1/2,\,1/5$, the radiation and the neutrinos behave exactly in the
same way as they do in the $\Lambda$CDM model so we expect that both
of these can reproduce the observed Universe.
The first values for $n$ are not able to explain the big bang nucleosynthesis,
since radiation never dominates as it is required. However, the last values
for $n$ can reproduce the radiation dominated era.
Following the radiation dominated era,
$\phi^2$ dark matter becomes the component that
dominates the evolution and finally the Universe is dominated by the
cosmological constant. Fig \ref{fig:Pe} shows the constraint $F$ in (\ref{eq:fri4})
in order to visualize the integration's error.
Observe that $F \approx 1$ at every point in the evolution, indicating
that the Friedmann equation is exactly fulfilled all the time,
this behavior is exactly the same for all
runs.

\subsection{The \textit{cosh} scalar potential}

Now, we are going to compare above results with the potential
\begin{equation}
V(\phi)=V_0\left[\cosh({\kappa}\lambda\phi)-1\right].
\label{eq:Vcosh}
\end{equation}

In order to do so, we define new variables as
\begin{eqnarray}
x& \equiv &\frac{\kappa}{\sqrt{6}}\frac{\dot\phi}{H},\nonumber \\
u& \equiv &\sqrt{\frac{2\,V_0}{3}}\frac{\kappa}{H}{{\cosh\left(\frac{1}{2}\kappa\lambda\phi\right)}},\nonumber \\
v& \equiv &\sqrt{\frac{2\,V_0}{3}}\frac{\kappa}{H}{{\sinh\left(\frac{1}{2}\kappa\lambda\phi\right)}},\nonumber\\
z_{\gamma}& \equiv &\frac{\kappa}{\sqrt{3}}\frac{\sqrt{\rho_{\gamma}}}{H},\,\,\,
l \equiv \frac{\kappa}{\sqrt{3}}\frac{\sqrt{\rho_\Lambda}}{H}.\label{eq:var2}
\end{eqnarray}

Substituting definitions (\ref{eq:var2}) into equations
(\ref{eq:ecua}) we obtain
\begin{eqnarray}
x'&=& -3\,x - \lambda v u+\frac{3}{2}\Pi\,x, \nonumber\\
u'&=& \lambda x v  + \frac{3}{2}\Pi\,u, \nonumber \\
v'&=& \lambda x u + \frac{3}{2}\Pi\,v, \nonumber \\
z_{\gamma}'&=&\frac{3}{2}\left(\Pi-\gamma \right)\,z_{\gamma}, \nonumber\\
l'&=&\frac{3}{2}\Pi\,l,  \label{eq:sd3}
\end{eqnarray}
where again the prime means derivatives with respect the e-folding
number $N=\ln(a)$ and we are also using the function
$\Pi=2x^2+\gamma z^2$. From the definitions (\ref{eq:var2}) it
follows the constraints
\begin{equation}
u^2-v^2=\frac{2V_0\,\kappa^2}{3}\frac{1}{H^2}=\frac{1}{\lambda^2}\frac{m_\phi^2}{H^2}
\label{eq:rest_cosh},
\end{equation}
and the Friedmann equation (\ref{eq:fri}) written in this variables
reads
\begin{equation}
F=x^2+u^2+z^2+l^2=1. \label{eq:Fridman}
\end{equation}

However, equation (\ref{eq:Fridman}) is actually not real
constraint, since they are inhered in the dynamical equations
(\ref{eq:sd3}) (see appendix A equation (\ref{eq:friGG})).
Furthermore, constraint (\ref{eq:rest_cosh}) is also inhered in
the dynamical system, observe that if we multiply the second
equation of (\ref{eq:sd3}) by $1/2\,u$ and the third by $1/2\,v$
and rest each other, we obtain
\begin{equation}
H'=-\frac{3}{2}\Pi\,H.
 \label{eq:constrain_Cosh}
\end{equation}

But this relation follows directly from the field equations
(\ref{eq:ecua}). This means that system (\ref{eq:sd3}) is compatible
with the constraint (\ref{eq:rest_cosh}).
Using this constraint (\ref{eq:rest_cosh}) in the dynamical system
(\ref{eq:sd3}), we obtain

\begin{eqnarray}
x'&=& -3\,x - u \sqrt{\lambda^2u^2+\left(\frac{m}{H}\right)^2}+\frac{3}{2}\Pi\,x, \nonumber\\
u'&=& x \sqrt{\lambda^2u^2+\left(\frac{m}{H}\right)^2} + \frac{3}{2}\Pi\,u, \nonumber \\
z'&=&\frac{3}{2}\left(\Pi-\gamma \right)\,z, \nonumber\\
l'&=&\frac{3}{2}\Pi\,l.  \label{eq:sd5}
\end{eqnarray}

We notice, that occurs the same situation as $\phi^{2}$ potential. Introducing
again the variable $s \equiv m/H$ with its dynamical equation.
\begin{equation}\label{eq:sss}
s'\,\,=\,\,\left(m t_0\right)^{\frac{1}{n}-1} n\,\,
\left(\frac{1}{s}\right)^{\frac{1}{n}-2},
\end{equation}
we obtain
\begin{eqnarray}
x'&=& -3\,x - u \sqrt{\lambda^2u^2+s^2}+\frac{3}{2}\Pi\,x, \nonumber\\
u'&=& x \sqrt{\lambda^2u^2+s^2} + \frac{3}{2}\Pi\,u, \nonumber \\
z_{\gamma}'&=&\frac{3}{2}\left(\Pi-\gamma \right)\,z_{\gamma},\nonumber\\
l'&=&\frac{3}{2}\Pi\,l, \nonumber\\
s'&=&s_0\,\,\left(\frac{1}{s}\right)^{\frac{1}{n}-2}.\label{eq:sd6}
\end{eqnarray}

\begin{figure}
\centerline{\epsfysize=5cm \epsfxsize=6.5cm \epsfbox{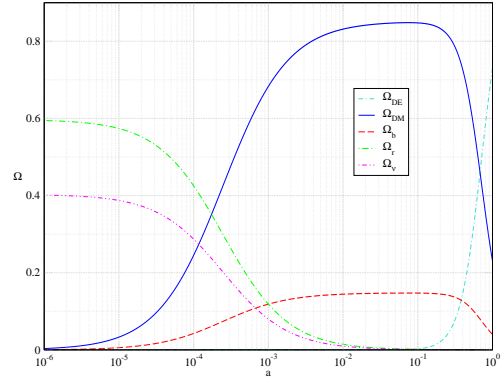}} \caption{ Evolution
of the density parameters for the system (\ref{eq:sd6}), where the
scalar field potential is given by the equation (\ref{eq:Vcosh}).}
\label{fig:cosh}
\end{figure}

The density parameters are the same as we have defined at
(\ref{eq:dens}). We solve numerically (\ref{eq:sd6}) with the same
initial conditions as the system of equations (\ref{eq:sd4}) and $\lambda
\approx 20$. The solutions are shown in Fig. (\ref{fig:cosh}).
The plot shows the dynamical evolution for a Universe with SFDM with
the potential (\ref{eq:Vcosh}), notice that is equivalent to potential
(\ref{eq:phi2}). 

Finally, we use the same dynamical system formalism for the
case of $\Lambda$CDM in order to compare with SFDM. We consider
that it background Universe is composed by baryons, radiation,
neutrinos, cold dark matter and cosmological constant with an
equation of state as perfect fluid. We solve numerically this system  
and in general terms the dynamic of both scalar potentials is indistinguishable
of the standard model. This is an important goal of this paper.

The next step is to compute the age of the Universe using our model.
The age equation can be written as
\begin{equation}
 t_{o}= \int_{N_o}^{N} \frac{1}{H}dN.
\label{eq:age}
\end{equation}

Using the definition for $l$ from (\ref{eq:var}) or (\ref{eq:var2}),
eq. (\ref{eq:age}) reduces to
\begin{equation}
 t_{o}= \frac{\sqrt{3}}{\kappa \sqrt{\rho_{\Lambda}}} \int_{N_o}^{N} l\,dN.
\label{eq:agef}
\end{equation}

We compute (\ref{eq:agef}) and obtain that $t_{o} \simeq 13.77$ Gyr.
This result is in agreement with the cosmological observations from
WMAP+BAO+SN which estimate $t_{o}=13.73 \pm 0.12$ Gyr and therefore
$H_{o}= 70.1 \pm 1.3$ $\textrm{km\,s}^{-1}\textrm{Mpc}^{-1}$. Furthermore, in Fig. \ref{fig:cosh},
we see that scale factor of decoupling is $a \sim 10^{-3}$, this
means a redshift $z \sim 1000$. At this redshift, the neutrinos made up
$\sim 12\%$ of the Universe. On the other hand,
WMAP cosmological observations show that when the Universe
was only 380,000 years old, neutrinos permeate the Universe within
$10\%$ of its total energy density.
Thus, SFDM is in agreement within the measurements of WMAP.
This result shows that scalar field is a plausible candidate
for dark matter because it behaves like cold dark matter.

\section{Conclusions}\label{sec:conclusiones}

SFDM has provided to be an alternative model for the dark matter
nature of the Universe. We have shown that the scalar field with a
ultralight mass condensates very early in the Universe and
generically form BEC's with a density profile which is very
similar as that of the CDM model, but with a almost flat central
density profile, as it seems to be in LSB and dwarf galaxies. This
fact can be a crucial difference between both models. If the flat
central density is no confirmed in galaxies, we can rule out the
SFDM model, but if this observation is confirmed, this can be a
point in favor of the SFDM model. We also show that the
$1/2m^2\phi^2$ potential and the $V_0[\cosh(\kappa\lambda\phi)-1]$
model are in fact the same. They have the same predictions, a
control variable which determines the behavior of the model, given
naturally the right expected cosmology and the same cosmology as
the CDM model. This implies that the differences between both
models, the CDM and SFDM ones, is in the non linear regime of
perturbations. In this way they form galaxies and galaxy clusters,
specially in the center of galaxies where the SFDM model predicts
a flat density profile. If the existence of supersymmetry is
confirmed, the DM supersymmetric particles would be observed by
detectors and they would have the right mass, DM density and coupling
constant, therefore the SFDM model can be ruled out. However, if these observations
are not confirmed, the SFDM is an excellent alternative candidate
to be the nature of the DM of the Universe.

\section*{Acknowledgements}

We would like to thank referee for his/her comments and 
Luis Ure\~na-L\'opez for reading the draft and his suggestions to improve the paper.
Also to Blanca Moreno Ley and J. Sanch\'ez-Salcedo for
many helpful and useful discussions. The numerical computations were
carried out in the "Laboratorio de Super-C\'omputo
Astrof\'{\i}sico (LaSumA) del Cinvestav" and 
in the "Super-Computadora KanBalam de la UNAM". 
This work was partly supported by CONACyT M\'exico, under grants 49865-F, 54576-F,
56159-F, and by grant number I0101/131/07 C-234/07, Instituto
Avanzado de Cosmologia (IAC) collaboration.

\appendix

\section[]{Dynamical system review.}\label{App:DynSys}

The theory of dynamical systems is used in the study of physical
systems that evolve over time. It is assumed that the physical state
of the system to an instant of time $t$ is described by an element
$x$ of a space phase $X$, which can be of finite or infinite
dimension. The evolution of the system is represented by a
differential autonomous equation in X, written symbolically as

\begin{equation}\label{eq:DE}
 \frac{d\mathbf{x}}{dt}=\mathbf{f}(\mathbf{x}), \hspace{1.5cm}
\mathbf{x}\,\, \epsilon \,\, X,
\end{equation}
where $\mathbf{f}: X \rightarrow X$.

The main step to get qualitative information on solutions is
studying the flow of the equation in the vicinity of their critical
points based on the Hartman-Grobman theorem, namely the study of its
stability.
\\
\\
The essential idea is firstly find the fixed (or critical) points of
the equation (\ref{eq:DE}) which are given by
$\mathbf{f}(\mathbf{x_c})=0$. Then linearized the differential
equation at each critical point, that is, expanding about these
points $\mathbf{\vec x}= \mathbf{\vec x_c} + \mathbf{\delta \vec x}$
which yields to
\[
\mathbf{\delta \vec x'} = \mathcal{M} \mathbf{\delta \vec x},
\]
where $\mathcal{M}$ is the Jacobian matrix of $\mathbf{\vec{x'}}$.
Therefore the general solutions for the linear perturbation
evolution can be written as
\[
\mathbf{\delta \vec x'}=\mathbf{\delta \vec x_0}e^{\mathcal{N}\delta
t},
\]
where $\mathcal{N}$ is the matrix composed of the eignvalues $m_i$
associated to $\mathcal{M}$.

The stability of the system (\ref{eq:DE}) depends on the values of
the eigenvalues: if the real part of all eigenvalues is negative,
the fixed point is asymptotically stable, i.e., an attractor. All
eigenvalues with positive real part make the fixed point
asymptotically unstable (commonly called as source or repeller).

On the other hand, a saddle point happens when there exists a
combinations of stable and unstable points. For a extended review
see, \citet{col}.
\\
Then, we give a procedure for transforming equations (\ref{eq:ecua})
and (\ref{eq:Frie}), with an arbitrary potential, into a dynamical
system. We define the dimensionless variables
\begin{eqnarray}
x &\equiv& \frac{\kappa}{\sqrt{6}}\frac{\dot\phi}{H},\,\,\,
u \equiv \frac{\kappa}{\sqrt{3}}\frac{\sqrt{V}}{H},\nonumber\\\,
z_{\gamma} &\equiv& \frac{\kappa}{\sqrt{3}}\frac{\sqrt{\rho_{\gamma}}}{H}.\,\,\,
\label{eq:varG}
\end{eqnarray}

Using above definitions (\ref{eq:varG}), the evolution
equations (\ref{eq:ecua}) transform into an autonomous system
\begin{subequations}
\begin{eqnarray}
x'&=& -3\,x +\frac{3}{2}\Pi\,x- \frac{\kappa}{\sqrt{6}\,H^2}
V_{,\phi}\label{eq:sdG_x},
\\
u'&=&  \frac{3}{2}\Pi\,u +
\frac{\kappa}{\sqrt{6}\,H^2}V_{,\phi}\,\frac{x}{u}\label{eq:sdG_u},
\\
z_{\gamma}'&=&\frac{3}{2}\left(\Pi-\gamma
\right)\,z_{\gamma}\label{eq:sdG_l},
\\
-\frac{ H'}{H}&=&\frac{3}{2}(2x^2+\gamma z_{\gamma}^2)\equiv
\frac{3}{2}\Pi  \label{eq:sdG_H}.
\end{eqnarray}\label{eq:sdG}
\end{subequations}

This last equation (\ref{eq:sdG_H}) can be written also as
\begin{equation}
s'=\frac{3}{2}\Pi\,s\label{eq:sdG_s},
\end{equation}
for the variable $s=cte./H$, and determines the evolution of the
horizon.
Here a prime denotes a derivative with respect to the e-folding
number $N=\ln(a)$. The Friedmann equation (\ref{eq:fri})
transforms  into a constraint equation
\begin{equation}\label{eq:friG}
F=x^2+u^2+z_{\gamma}^2=1.
\end{equation}
With these variables, the SFDM density can be written as
\begin{equation}\label{eq:densG}
\Omega_{DM}=x^2+u^2. \nonumber
\end{equation}

Observe that if we derive (\ref{eq:friG}) with respect to $N$ and
substitute system (\ref{eq:sdG}) into this, we obtain
\begin{equation}\label{eq:friGG}
F'=3\,(F-1)\,\Pi,
\end{equation}
indicating that constraint (\ref{eq:friG}) is compatible with
system (\ref{eq:sdG}) for all scalar field potentials if the
Friedmann equation is fulfilled.

Now we show that system (\ref{eq:sdG}) together with constraint
(\ref{eq:friG}) is completely integrable. To integrate system
(\ref{eq:sdG}), first observe that we can substitute $3/2\Pi$ from
equation (\ref{eq:sdG_s}) into the rest of the equations. With
this substitution equation (\ref{eq:sdG_l}) can be integrated in
terms of $s$ as
\begin{equation}
z_{\gamma}=\sqrt{\Omega{^{(0)}_{\gamma}}}\,s\,\exp(-\frac{3}{2}\gamma\,N)\label{eq:lfinalG},
\end{equation}
where $\Omega{^{(0)}_{\gamma}}$ is an integrations constant. Now we
multiply (\ref{eq:sdG_x}) by $2\,x$ and (\ref{eq:sdG_u}) by $2\,u$
and sum both equations. We obtain
\begin{equation}
(x^2+u^2)'=-6\,x^2 + 2\,\ln(s)'(x^2+u^2)\label{eq:x2+u2G}.
\end{equation}

Now, we use constraint (\ref{eq:friG}) and equation
(\ref{eq:lfinalG}) into equation (\ref{eq:x2+u2G}) to obtain
\begin{equation}
6\,x^2=2\,\ln(s)'-3\gamma\,s^2\,\Omega{^{(0)}_{\gamma}}\exp(-3\,\gamma\,N)\label{eq:x2G}.
\end{equation}

Now we have to integrate equation (\ref{eq:sdG_s}) with all
these results. If we substitute (\ref{eq:x2G}) and
(\ref{eq:lfinalG}) into (\ref{eq:sdG_H}) or (\ref{eq:sdG_s}) we
obtain $0=0$, that means $s$ is an arbitrary variable 
which parametrizes the decrease of H
and can be cast into the system as a control variable. In other words, equations
(\ref{eq:sdG_H}) and (\ref{eq:sdG_s}) are actually identities, and
not equations.

Thus, we set the variable $s$ from system (\ref{eq:sdG}) arbitrary
in the equations (\ref{eq:sdG_x}), (\ref{eq:sdG_u}) and
(\ref{eq:sdG_l}).


\label{lastpage}
\end{document}